\documentclass[superscriptaddress,twocolumn,aps,prc,showpacs,10pt]{revtex4}

\usepackage{amsmath}
\usepackage{amssymb}
\usepackage{graphicx}
\usepackage{mathrsfs}
\renewcommand{\vec}[1]{\mbox{\boldmath $#1$}}

\begin{document}
\title{Application of the inverse Hamiltonian method to Hartree-Fock-Bogoliubov calculations}

\author{Y. Tanimura}
\affiliation{Department of Physics, Tohoku University, Sendai 980-8578, Japan}
\author{K. Hagino}
\affiliation{Department of Physics, Tohoku University, Sendai 980-8578, Japan}
\author{P. Ring}
\affiliation{Physik Department, Technische Universit\"at M\"unchen, D-85748 Garching, Germany}

\begin{abstract}
We solve the Hartree-Fock-Bogoliubov (HFB) equations for a spherical mean field 
and a pairing potential with the inverse Hamiltonian method, which we have developed for the 
solution of the Dirac equation.
This method is based on the variational principle for the inverse Hamiltonian,
and is applicable to Hamiltonians that are bound neither from above nor below.
We demonstrate that the method works well not only for the Dirac but also
for the HFB equations.
\end{abstract}


\maketitle

Pairing correlations between nucleons play a very important role in open shell nuclei \cite{Brink,RS}.
Hartree-Fock-Bogoliubov (HFB) theory is a powerful method which treats these correlations in a
self-consistent way in the framework of a single generalized Slater determinant of independent
quasi-particles \cite{RS,Bender,Doba}. The method has been widely used in recent years for the
study of the structure of neutron rich nuclei far from stability up to the neutron drip line,
where the coupling to the continuum has a great influence.

\begin{figure}
\begin{center}
\includegraphics[scale=.7]{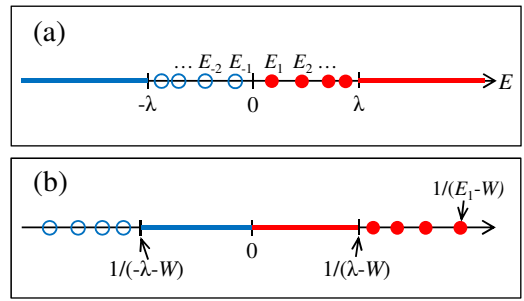}
\end{center}
\caption{(Color Online) Spectra of (a) a quasi-particle Hamiltonian $H$ itself and (b) the inverse
of the Hamiltonian $1/(H-W)$. $\lambda$ is the chemical potential. The bound states of positive and
negative energies are indicated by solid and open circles, respectively.
The continuum states are represented by the thick solid lines.
The energy shift $W$ is taken between the positive and negative spectra.
The eigenvalues are labeled by an integer $k$ such that $E_{-k}=-E_k$. }
\label{fig:spectrum}
\end{figure}

The HFB-equations are a set of coupled differential equations. In many cases they have been solved by
an expansion in terms of a finite set of basis functions, as for instance the eigenfunctions of a
harmonic oscillator or a Woods-Saxon potential. Although this method has been used successfully for
many investigations in the literature it has its limitations.
(i) The convergence with the number of basis functions depends on the parameters of the basis and the optimization of this parameters is often very complicated, (ii) in many cases, in particular for two-dimensional (2D) and three-dimensional (3D) calculations in heavy nuclei, the dimension of the matrices becomes extremely large. Since the CPU time for one diagonalization grows with the cube of this dimension this is connected with considerable numerical efforts, (iii) In each step of the iteration the corresponding matrix elements of the two-body interaction in terms of the basis functions requires in addition much CPU time, (iv) the treatment of the continuum is connected with specific difficulties in particular in the case of neutron halos. In summary these methods do not exploit fully the advantages of powerful zero-range interactions as for instance the Skyrme energy density functional. These advantages can only be exploited fully in coordinate space. For this reason the
  Orsay group has introduced already in the eighties for the solution of the Hartree Fock (HF) equations the imaginary time method \cite{Davies1980_NPA342-111,EV8}. Starting with an initial HF wave functions $|\Psi^{(0)}\rangle$ the solution is obtained in iterative steps with infinitesimal step size $\tau$ as
\begin{equation}
|\Psi^{(n+1)}\rangle\propto e^{-\tau \hat{H}}|\Psi^{(n)}\rangle
\end{equation}
Obviously this method is limited to Hamiltonians $H$ with a spectrum bounded from below, as the non-relativistic HF equation. An important fact is that the spectrum of the HFB equation is bound neither from above nor from below (see Fig. \ref{fig:spectrum}(a)) because of the coupling between particle creation and annihilation parts in the quasi-particle operators of the Bogoliubov transformation \cite{RS}. This inhibits a direct application of the imaginary time method to HFB, which has been successfully employed in self-consistent mean field calculations in the coordinate space representation \cite{Bonche85,EV8}. That is, if the imaginary time evolution is naively applied, the iterative solution
inevitably dives into the quasi-particle negative continuum. To avoid this problem the rather complicated two-basis method has been introduced in Refs. \cite{Bonche85,Terasaki96} where, in each step of the iteration, the HFB equations are solved by expansion of the quasi-particle wave functions in a HF-basis calculated by the imaginary time method on a 3D mesh in coordinate-space. In Ref. \cite{Tajima04} the HFB equation on a 3D mesh has been solved in the canonical basis.

It should be mentioned that the diagonalization of the huge matrices in the basis expansion method can be avoided by the gradient method introduced in Ref. \cite{Mang1976_ZPA279-325}, which is also applicable to the solution of the HFB equation with a spectrum not bound from below. Here the wave function in the next step of the iteration is expressed in terms of the Thouless theorem \cite{RS}
\begin{equation}
|\Psi^{(n+1)}\rangle\propto \exp\left(-\tau\sum_{k<k'}H^{20}_{kk'}\alpha^\dag_{k^{}}\alpha^\dag_{k'}\right)|\Psi^{(n)}\rangle
\end{equation}
where $|\Psi^{(n)}\rangle$ is the vacuum with respect to the quasi-particle operators $\alpha^\dag_k$. For infinitesimal $\tau$ only 2-quasi-particle states with positive energy $E_k+E_{k'}$ are admixed in the next step of the iteration. This method has been successfully used
in the literature for the solution of the HFB equations in an oscillator basis \cite{Egido1995_NPA594-70} and it has been also applied for a variation after
projection (VAP) in Refs. \cite{Egido1982_NPA388-19,Anguiano2001_NPA696-467}. Of course, the matrix elements $H^{20}_{kk'}$ have to be calculated here in the corresponding quasi-particle basis. Therefore the locality of the Skyrme interaction and the quasi-locality of the kinetic energy in coordinate space cannot be exploited in this method.

We therefore propose in this report for the solution of the HFB equations a different method keeping in mind that one is confronted with the same problem in the solution of the Dirac equation because the corresponding spectrum has the Dirac sea states down to the negative infinity as well as the positive energy states up to the positive infinity. This leads to a breakdown of variational calculations which has long been known in the field of relativistic quantum chemistry under the name of ``variational collapse'', and there has been a number of prescriptions proposed to avoid it \cite{WaKu81,WaKu83,StHa,HiKr,FaFoCh}. Recently, newly developed methods for iterative solutions of a Dirac equation are
introduced by Zhang {\it et al.} \cite{ZhLiMe,ZhLiMe-f} and by Hagino and Tanimura \cite{HaTa10} in the nuclear physics context.
In Ref. \cite{HaTa10}, based on the idea of Hill and Krauthauser \cite{HiKr} a novel method has been developed, which is called ``inverse Hamiltonian method'', for relativistic
mean field calculations in the coordinate space representation.
In this method a variational principle is applied to the Hermitian operator $1/(H_{\rm Dirac}-W)$
instead of the Hamiltonian $H_{\rm Dirac}$ itself.
Here, $W$ is a real number which is set between the Fermi sea and the Dirac sea.
In contrast to some other methods \cite{WaKu83,StHa,ZhLiMe,ZhLiMe-f},
it is relatively straightforward to apply our method
not only to the Dirac equation but also to other eigenvalue problems with 
unbound operators, such as HFB. 
In this paper, we apply the inverse Hamiltonian method to a HFB equation in the
coordinate space representation and
show that the equation can be solved successfully without the variational collapse.

In HFB calculations one usually needs to obtain several lowest positive energy quasi-particle
states only. That is, one only needs the states $\psi_1, \psi_2, ...$ associated with
eigenvalues $E_1, E_2, ...$ (see Fig. 1(a))
As is seen in Fig. \ref{fig:spectrum}(b), these states come to the top of the spectrum
of $1/(H-W)$ if $W$ is set between the positive and negative spectra.
A variational principle for $1/(H-W)$ is that a maximization of $\langle (H-W)^{-1}\rangle$
leads to the desired quasi-particle state wave functions \cite{HiKr}.
Our method maximizes $\langle (H-W)^{-1}\rangle$ based on the relation \cite{HaTa10}
\begin{equation}
|\psi^{(n+1)}\rangle\propto \exp\left(\frac{\Delta T}{H-W}\right)|\psi^{(n)}\rangle
\end{equation}
where $|\psi^{(0)}\rangle$ is an arbitrary wave function which is not an eigenfunction
of the Hamiltonian $H$ and $W$ is a real constant between $E_1$ and $E_{-1}$.
All the states below $W$ damp out and only $\psi_1$ which is just above $W$ survives
in the limit $n\to\infty$.

In practice, the wave function 
$\psi^{(n+1)}$ is evolved for a small step $\Delta T$ from $\psi^{(n)}$ as 
\begin{equation}
|\psi^{(n+1)}\rangle\propto\left(1+\frac{\Delta T}{H-W}\right)|\psi^{(n)}\rangle=|\psi^{(n)}\rangle + \Delta T|\phi^{(n)}\rangle
\end{equation}
To this end, we have to solve a large sparse linear equation
\begin{equation}
(H-W)|\phi^{(n)}\rangle=|\psi^{(n)}\rangle
\label{eq:inv}
\end{equation}
in order to invert the Hamiltonian.
We here employ an iterative method for linear systems, that is, the conjugate gradient normal
residual (CGNR) method \cite{Saad}. This is one of the Krylov subspace
methods for sparse linear systems \cite{Vorst,Saad}.
CGNR solves a linear system $A\vec{x}=\vec{b}$ by applying the conjugate
gradient method to an equivalent system $A^{\dagger}A\vec{x}=A^{\dagger}\vec{b}$.

When the mean field and pairing potentials are local and spherical, a quasi-particle wave
function is given by the form
\begin{equation}
\psi_k(\vec{r})=
\frac{1}{r}\left(\begin{array}{c}
U_k(r)\mathscr{Y}_{\ell jm}(\theta,\phi)\\
V_k(r)\mathscr{Y}_{\ell jm}(\theta,\phi)
\end{array}\right).
\end{equation}
Here, $\mathscr{Y}_{\ell jm}$ is a spherical spinor defined by
$\mathscr{Y}_{\ell jm}=\sum_{m,m'}\langle\ell m\frac{1}{2}m'|jm\rangle Y_{\ell m}\chi_{m'}$
, where $Y_{\ell m}$ and $\chi_{m'}$ are spherical harmonics and spin wave function, respectively.
The HFB equation in the coordinate space then reduces to a the radial equation
\begin{equation}
\left(\begin{array}{cc}
h-\lambda & \Delta(r) \\
\Delta(r) & -h+\lambda
\end{array}\right)
\left(\begin{array}{c}
U_k(r)\\
V_k(r)
\end{array}\right)
=
E_k\left(\begin{array}{c}
U_k(r)\\
V_k(r)
\end{array}\right),
\label{eq:HFBeq}
\end{equation}
where $h$ is the mean field Hamiltonian, $\lambda$ is the chemical potential, and
$\Delta(r)$ is the pairing potential.
Following Refs. \cite{HaMo03,HaSa05}, we use a phenomenological Woods-Saxon type
potentials, which simulates medium-heavy neutron-rich nuclei around $^{84}$Ni,
for the mean field and the pairing potentials.
The potential $v(r)$ in the mean field Hamiltonian $h$ and the paring potential $\Delta(r)$
are thus taken as
\begin{eqnarray}
v(r)&=&v_0f(r)+v_{\ell s}\frac{1}{r}\frac{df}{dr}\vec{\ell}\cdot\vec{s}, \\
\Delta(r)&=&\Delta_0f(r), \\
f(r)&=&\frac{1}{1+e^{(r-R_0)/a}},
\end{eqnarray}
with $v_0=-38.5$ MeV,  $v_{\ell s}=14$ MeV$\cdot$fm$^{2}$, $R_0=5.63$ fm,
and $a=0.66$ fm \cite{HaMo03,HaSa05}.
The strength of pairing potential $\Delta_0$ is determined so that the average pairing
gap $\bar\Delta$ defined by \cite{HaMo03}
\begin{equation}
\bar\Delta=\frac{\int_0^{\infty}r^2dr\ \Delta(r)f(r)}{\int_0^{\infty}r^2dr\ f(r)}
\end{equation}
is equal to 1.0 MeV.
The chemical potential $\lambda$ is fixed to $\lambda=-0.5$ MeV in the present calculation.
We solve Eq. (\ref{eq:HFBeq}) by discretizing the radial coordinate $r$ with mesh size $\Delta r$,
and imposing the box boundary condition. The second derivative of $\psi$ at the $i$th
mesh point is approximated by 3-point difference: $\psi''_i=(\psi_{i+1}-2\psi_i+\psi_{i-1})/(\Delta r)^2$.

Let us now apply the inverse Hamiltonian method and numerically solve the HFB equation,
Eq. (\ref{eq:HFBeq}).
We also solve the equation exactly by directly diagonalizing the coordinate space Hamiltonian.
by the Runge-Kutta method. The parameters of the inverse Hamiltonian method are set $W=0.1$ MeV and $\Delta T=10$ MeV.
The excited states are also calculated simultaneously by orthogonalizing a set of
wave functions at every step of iteration.
The radial coordinate is discretized up to $r_{\rm max}=30$ fm with $\Delta r=0.1$ fm.
Initial quasi-particle wave functions are taken to be a Gaussian form
\begin{equation}
\left(\begin{array}{c}
U_k^{(0)}(r)\\
V_k^{(0)}(r)
\end{array}\right)
=N_k\left(\begin{array}{c}
r^{\ell+1}e^{-r^2/b_k^2} \\
r^{\ell+1}e^{-r^2/b_k^2}
\end{array}\right),
\end{equation}
where $\ell$ is the orbital angular momentum and $N_k$ is an appropriate normalizing constant.
The width parameter of the Gaussian $b_k$ is taken as $b_k=2.5\times1.05^{k-1}$ fm, $(k=1,2,...)$.

\begin{figure}
\vspace{.6cm}
\begin{center}
\includegraphics[scale=.4,angle=-90]{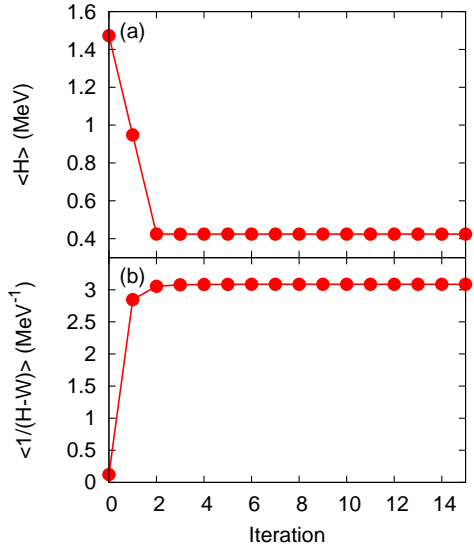}
\end{center}
\caption{(Color Online) Covergence properties of (a) the energy expectation value
$\langle H\rangle$ and (b) the expectation value of the inverse of Hamiltonian
$\langle (H-W)^{-1} \rangle$ for the lowest $s_{1/2}$ quasi-particle state.
The energy shift and the step size of $T$ are taken to be $W=0.1$ MeV and $\Delta T=10$
MeV, respectively.}
\label{fig:econv}
\end{figure}

Let us first discuss the convergence properties of the energy $\langle H\rangle$ and
the expectation value of the inverse of Hamiltonian $\langle (H-W)^{-1} \rangle$ for
the lowest $s_{1/2}$ quasi-particle state.
In Fig. \ref{fig:econv}, we show the evolution of the two quantities as functions of
the number of iteration steps.
As is observed in Ref. \cite{HaTa10} for a Dirac equation, $\langle (H-W)^{-1} \rangle$
converges monotonically up to a certain value as the iteration step increases.
At the same time, $\langle H\rangle$ converges to the lowest $s_{1/2}$ eigenvalue, $E=0.424$ MeV.

\begin{figure}
\begin{center}
\hspace{.5cm}
\includegraphics[scale=.45,angle=-90]{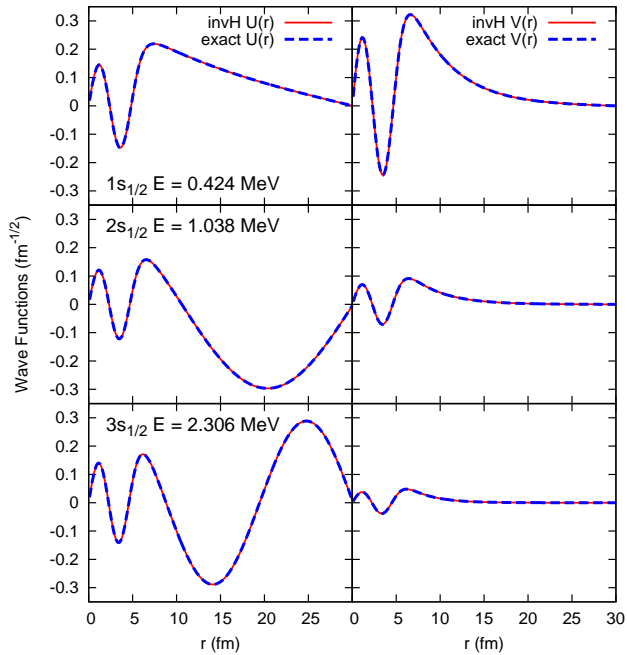}
\end{center}
\caption{(Color Online) Comparisons of wave functions for the three lowest $s_{1/2}$ states.
The left and right panels show the upper ($U_k(r)$) and the lower ($V_k(r)$) components of
quasi-particle wave function, respectively. The exact wave functions are
shown with the dashed lines and the ones obtained by the inverse Hamiltonian method are
drawn with the solid lines. }
\label{fig:wf}
\end{figure}

In Table \ref{tb:Eqp}, we show quasi-particle energies and occupation probabilities $v_k^2$ for the
three lowest $s_{1/2}$ states
in comparison with the exact values which are obtained by diagonalizing the Hamiltonian.
The occupation probabilities are defined in terms of quasi-particle wave function by
\begin{equation}
v_k^2=\int_0^{\infty}dr\ |V_k(r)|^2.
\end{equation}
The agreement is perfect both in the energies and the occupation probabilities
for the digits shown in the table.
Fig. \ref{fig:wf} shows comparisons of wave functions of the three $s_{1/2}$ states.
The dashed lines show the exact wave functions, whereas the solid lines
show the wave functions obtained with the inverse Hamiltonian method.
The left and right panels show the upper component $U_k(r)$ and the lower component $V_k(r)$
of a quasi-particle wave function, respectively.
As is seen in Fig \ref{fig:wf}, the inverse Hamiltonian method reproduces the wave functions
almost identically to the exact ones for both the bound state and the excited continuum states.
We have also obtained the other $s$-wave states with an accuracy
as high as the lower states shown in Table \ref{tb:Eqp} and Fig. \ref{fig:wf}.

We have checked the performance of the inverse Hamiltonian method for other angular
momentum quantum numbers and confirmed that the method solves the HFB
equation as accurately as for the $s_{1/2}$ states.
It is apparent that the inverse Hamiltonian method gives practically the exact solutions
of the HFB equation in the coordinate space representation and is safe against the variational collapse.

\begin{table}
\caption{A comparison between the exact calculations and the inverse Hamiltonian method for the
three lowest $s_{1/2}$ quasi-particle energies $E$ and occupation probabilities
$v_k^2$. The exact values are calculated by diagonalizing the real space Hamiltonian. }
\begin{center}
\begin{tabular}{cc|cc}
\hline\hline
\multicolumn{2}{c|}{$E$ (MeV)} & \multicolumn{2}{c}{$v_k^2$}\\
exact & inv. H method & exact & inv. H method\\
\hline
0.42414 & 0.42414 & 0.5574 & 0.5574\\
1.0383 & 1.0383 & $3.972\times 10^{-2}$ & $3.972\times 10^{-2}$\\
2.3063 & 2.3063 & $9.689\times 10^{-3}$ & $9.689\times 10^{-3}$\\
\hline\hline
\end{tabular}
\end{center}
\label{tb:Eqp}
\end{table}

In summary, we have discussed the numerical performance of the inverse
Hamiltonian method for a HFB calculation.
While the method has been developed for solving Dirac equations, we have shown that
it can almost exactly solve a coordinate space HFB equation as well with spherical
mean field and pairing potentials without variational collapse.
An obvious future work is the application of the method to self-consistent
HFB calculations on 3D mesh.  A work in this direction is now in progress.

This work was supported by Grant-in-Aid for JSPS Fellows under the program number
24$\cdot$3429 and the Japanese Ministry of Education, Culture, Sports, Science and
Technology by Grant-in-Aid for Scientific Research under the program number (C) 22540262.
The work of Y. T. was also supported by the Japan Society for Promotion of Science for
Young Scientists.

\end{document}